# Imaging Temperature and Thickness of Thin Planar Liquid Water Jets in Vacuum

Tillmann Buttersack, Henrik Haak, Hendrik Bluhm, Uwe Hergenhahn, Gerard Meijer, and Bernd Winter

Fritz-Haber-Institut der Max-Planck-Gesellschaft, Faradayweg 4–6, 14195 Berlin, Germany

Authors to whom correspondence should be addressed: buttersack@fhi-berlin.mpg.de, winter@fhi-berlin.mpg.de

**Abstract**

We present spatially resolved measurements of the temperature of a flat liquid water microjet for varying pressures, from vacuum to 100% relative humidity. The entire jet surface is probed in a single shot by a high-resolution infrared camera. Obtained 2D images are substantially influenced by the temperature of the apparatus on the opposite side of the IR camera; a protocol to correct for the thermal background radiation is presented. In vacuum, we observe cooling rates due to water evaporation on the order of $10^5$ K/s. For our system, this corresponds to a temperature decrease of approximately 15 K between upstream and downstream positions of the flowing leaf. Making reasonable assumptions on the absorption of the thermal background radiation in the flatjet we can extend our analysis to infer a thickness map. For a reference system our value for the thickness is in good agreement with the one reported from white light interferometry.

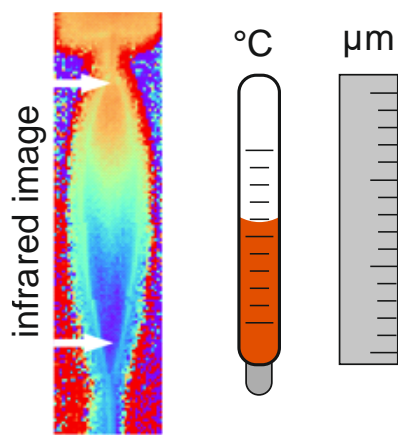

*TOC graphic*

**Introduction**

Fast-flowing liquid microjets (LJ) in vacuum are excellently suited for the study of the bulk and surface properties of aqueous and organic solutions using electron (*1-5*) and optical spectroscopy as well as X-ray absorption and emission spectroscopy.(*6-9*) Most commonly, cylindrical jets were utilized. More recently vacuum flatjets (FJ, also referred to as liquid sheets) with a planar surface have attracted considerable attention, and different techniques for their generation have been developed. These include the collision of two cylindrical LJs, application of asymmetric gas pressure on a cylindrical jet, the use of slit nozzles, or fan spray nozzles.(*8, 10-27*) One major advantage of FJs over cylindrical LJs is that the planar surface is the more suitable geometry for the quantitative interpretation of molecular beam scattering off the liquid surface, (*28*) and this also applies for quantitative measurements of the angular distributions of photoelectrons.(*29*) Furthermore, the planar surface is favorable for time-resolved photoelectron spectroscopy experiments, and enables the measurement of undistorted X-ray transmission spectra for sufficiently thin FJs. (*8*)

Figure 1a shows a sketch of a typical FJ, generated by two colliding cylindrical jets. In fact, upon impact a chain of connected ellipsoidal thin mutually orthogonal leaves separated by thicker nodes forms; Figure 1a looks in the direction of the surface normal of the second leaf. Each leaf is bound by a thicker fluid rim.(*12*) These thin leaves are however not truly planar but rather exhibit slightly curved surfaces, as has been observed experimentally and in theoretical modelling. (*10, 30, 31*) The (local) thickness of a FJ can be readily determined by the absorption of monochromatic light (e.g., in the infrared (*8, 17*) or X-Ray (*8, 14*) region of wavelengths). Many such measurements, at different positions on the leaf, one after the other, would be required to coarsely map the thickness distribution over a given FJ leaf. Alternatively, the relative thickness gradient across the leaf can be evaluated from interference patterns produced by white or monochromatic light.(*8, 17, 18, 32*) With a reference measurement the 4thickness of the leaf can then be spatially resolved.(*18*) These methods have been applied in several studies, using different designs for FJ generation, resulting in a large range of thicknesses at the center, ranging from a few tens of micrometers down into the sub-micron range.(*8, 13, 14, 17, 18, 20, 23, 33*)

Temperature measurements from LJ in vacuum have not been routinely performed and have prohibited to accurately access temperature-dependent properties from (aqueous) solutions, in particular in conjunction with electron spectroscopy. One desirable goal would be the determination of enthalpies or entropies, associated with interfacial (as opposed to bulk solution) chemical equilibria. Another challenge is the quantitative determination of metastable (supercooled) solution phases. Somewhat related, several studies and simulations already demonstrated a significant effect of the temperature on the hydration shell and on the solvent–solvent as well as solvent–solute interactions in aqueous solutions.(*34-39*) On a more practical note, knowledge of the accurate temperature is also essential when comparing measurements performed in different laboratories.

LJs in vacuum effectively cool by molecular evaporation, leading to a temperature gradient along the propagation direction.(*40*) However, the direct measurement of the temperature of thin jets is experimentally challenging. Contact techniques, *e.g.*, the use of tiny thermocouples, inevitably distorts the LJ, and even the FJ (although temperatures can be measured (*25*)) Cooling rates from a FJ, of similar size as in the present study, have been estimated to about $1.9 \cdot 10^5$ K/s, based on the analysis of measured mass loss (~5%).(*8, 26*) This approach however lacks information about local temperatures. An early indirect temperature measurement of cylindrical jets by Faubel *et al.* was based on the measured velocity distribution of evaporating water molecules.(*40*) Fitting a modified Maxwell-Boltzmann distribution revealed absolute local jet temperature (as low as 210 K) and cooling rates of $1.7 \cdot 10^5$ K/s.(*40*) These experiments indicate that deeply super-cooled and even hyper-cooled water (*41*) might be generated by evaporative cooling in LJs. Very recently, Chin *et al.* measured the respective velocity distributions to characterize the evaporation and molecular beam scattering from dodecane and neon-doped dodecane flat liquid jets.(*28*) Furthermore, Raman spectroscopy has been applied to evaluate the temperature of water both from cylindrical LJs (*42*) and micrometer-sized droplets.(*43-46*) Later, Nunes *et al.* applied static diffraction to determine the temperature of water FJs and reported cooling rates of up to $10^6$ K/s.(*47*) Recently, Chang *et al.* investigated the effect of the nozzle geometry and solvent on the temperature of the FJ, also by Raman spectroscopy.(*48*) The necessity to perform the measurement for each surface point of interest, one-by-one, is a considerable disadvantage of above method.

In the present study, we describe an approach, using an infra-red (IR) camera, to monitor the temperature as well as the thickness of a water FJ with a precision of ±1 K, simultaneously of the entire surface, with tens of μm-range spatial resolution. Previously, IR cameras have been applied to monitor, *e.g.*, the surface temperature of small freezing water droplets.(*49*) In the case of FJs, the temperature measurement *via* an IR camera is complicated by the fact that the liquid leaves are sufficiently thin so that they partially transmit the IR radiation from the background, with the transmission being dependent on the local jet thickness. We describe a protocol enabling the determination of the FJ-position-dependent thickness $d(x,z)$ and temperature $T(x,z)$, and associated evaporation rates, for different pressures of the atmosphere surrounding the jet, based on IR camera images.

**Methods**

We use two cylindrical water LJs, each with a diameter of 64 μm, and a combined flow rate of 6.2 mL/min (2x 3.1 mL/min), colliding at an angle of 45°, to create a FJ; compare Figure 1a. The velocity of the impinging LJs is approximately 16 m s$^{-1}$. At the point of injection, the jets are at room temperature. For details on the sample delivery system, we refer to Refs. (*33, 50*). In the center of the leaf (along the z-axis; see Fig. 1a) the velocity increases by about a factor of 1.3 with respect to that of the impinging cylindrical jets, amounting to 21 m s$^{-1}$; (*51*) the surface area of the leaf is approximately 3 x 0.7 mm$^2$. This information enables to estimate cooling rates.

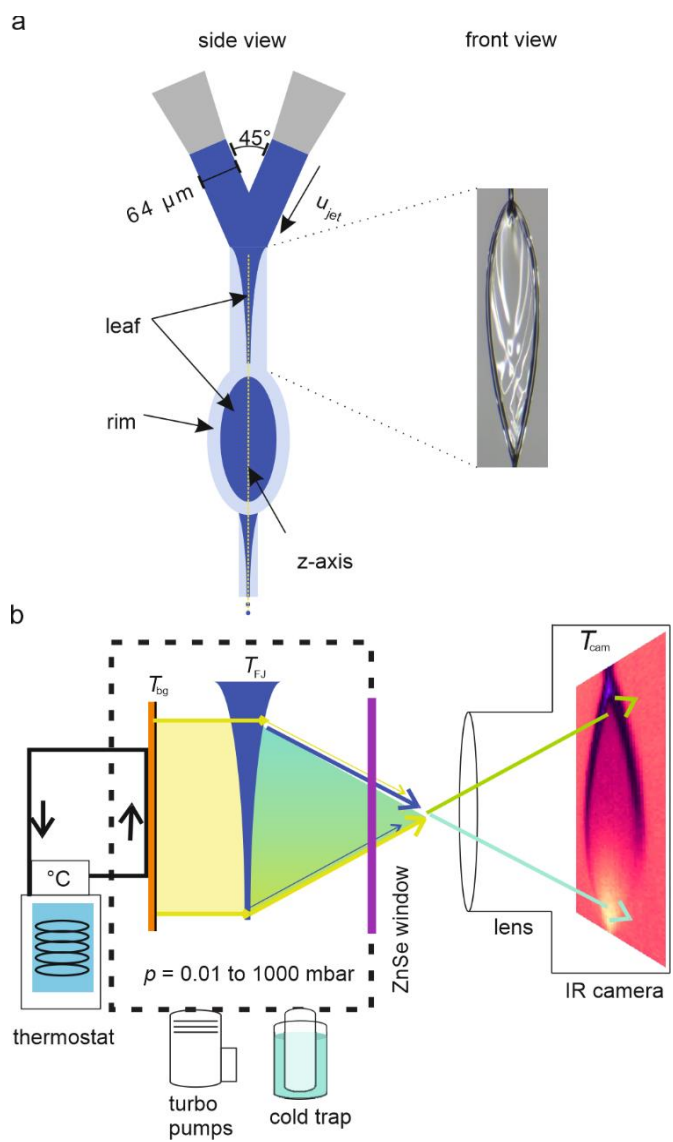

*Figure 1*: *Sketch of the experimental setup for measuring a 2D-map of the temperature and the thickness of a thin liquid film (flatjet, FJ) in a vacuum chamber. A (left): Two impinging cylindrical microjets form a chain of several leaves. Each leaf consists of a rim (light blue) surrounding the thin sheet (dark blue). First and third leaves are seen from the side, while the second leave is in paper plane. A (right): Actual photograph (as seen by eye) from the first leaf, now rotated by 90° along the jet-flow axis. B: The FJ is monitored with an infrared camera through an infrared-transmissive ZnSe window (purple). The temperature of the background behind the FJ (copper plate coated with black plastic tape to have a similar IR emission coefficient than water, orange) can be adjusted with a thermostat. The IR radiation from the background ($T_{bg}$, yellow) is partially transmitted through the leaf. The infrared camera records a temperature $T_{cam}$ (turquoise to green) which is in between the temperature of the background*

*and the FJ. By varying the background temperature, the position-dependent contribution of the background emission relative to the emission from the FJ can be extracted.*

The vacuum chamber is equipped with two roughing pumps and two turbomolecular pumps as well as two liquid-nitrogen cold traps. Under FJ operation conditions, and with all pumps in operation, the pressure in the vacuum chamber is in the low-$10^{-2}$ mbar range. We do report though on measurements at higher pressures as well, maintained when operating the mechanical pumps at reduced power, or completely switched off. In the latter case a open 500-mL water reservoir was placed in the chamber, which was backfilled with nitrogen to 1000 mbar (100% RH). To achieve slightly reduced RH pumps are switched on for a few seconds until the desired pressure is reached and then switched off. Note, if pumps are operating vapor is constantly removed (steady state), while this is not the case when then pumps are switched off.

Figure 1b shows a schematic of the overall experimental setup; the first leaf is shown from the side (in blue). The vacuum chamber is equipped with an IR-transmissive window (zinc selenide with anti-reflection coating, Artifex, in purple). Outside the chamber, behind the window, at a distance of approximately 80 mm from the surface of the water leaf, the IR camera (Optris PI640, software: PIX connect, 32 Hz) is positioned. Using a macro lens to image the transmitted IR light onto the camera yields to the detection of ~2000 data points of the leaf (28 µm/pixel spatial resolution) in a single snapshot, acquired in less than a second, enabling determination of thickness and temperature of the FJ with high spatial resolution. On the other side of the leaf, *i.e.*, facing away from the camera, a polyethylene film-covered (black duct tape, *Tesa*) copper plate (5x5 cm$^2$; in orange) is placed inside the vacuum chamber. Confirmed by calibration of the camera-determined plate temperature with the one measured using a thermocouple, the background (plate) has a very similar emission coefficient as water ($0.95 \approx \varepsilon_{bg} \approx \varepsilon_{FJ}$).(*52, 53*) Important for the present study, the plate temperature can be controlled between approximately 10 to 50 °C with a thermostat (Julabo, 300F). The temperature of the background $T_{bg}$ and temperature of the FJ $T_{FJ}$ are simultaneously recorded, as described in the SI (see Figure S3).

As we will explain, the as-measured temperature, $T_{cam}$, must be corrected at each surface point by a factor to account for the respective local thickness. We note that an approximately 100-μm thick water film would absorb the infrared radiation completely.(54) However, our FJ is significantly thinner at all positions, implying that IR radiation is partially transmitted.

The IR camera records the integrated power ($P$) within its sensitive wavelength region, 8—14 μm, and calculates from $P$, using the Stefan-Boltzmann equation ($P = A \cdot \sigma \cdot T^4$, $A$: area, $\sigma$: Stefan-Boltzmann constant) the temperature $T$ of an object. However, a distinction cannot be made if the radiation is emitted from the background (bg) or the liquid FJ since the power recorded by the camera $P_{cam}$ is the sum of the two contributions:

$$P_{cam}(x,z) = P_{FJ}(x,z) + P_{bg}(x,z) \qquad (1)$$

$T_{cam}(x,z)$ will be in between the actual temperature of the FJ, $T_{FJ}(x,z)$ and the temperature of the background, $T_{bg}(x,z)$, and can generally be expressed as:

$$T^4_{cam}(x,z) = T^4_{FJ}(x,z) \cdot + \beta(x,z) \cdot \left(T^4_{bg}(x,z) - T^4_{FJ}(x,z)\right) \qquad (2)$$

for temperatures within a small interval around $T_{FJ}$ and $T_{bg}$. Here, $\beta(x,z)$ is a correction matrix (with $0 \leq \beta(x,z) \leq 1$). It can be determined from linear regression when plotting $T^4_{cam}(x,z)$ *versus* $T^4_{bg}(x,z)$ for each pixel of the leaf ($x$, $z$), and necessitates recording $T_{cam}$ for a series of $T_{bg}$ values. Rewriting eq. (2) we obtain the following expression for the temperature of the FJ:

$$T_{FJ}(x,z) = \left(\frac{T^4_{cam}(x,z) - \beta(x,z) \cdot T^4_{bg}(x,z)}{(1-\beta(x,z))}\right)^{0.25} \qquad (3)$$

**Results**

In a first step, we evaluated the correction matrix *β(x,z)* for the FJ at 100% relative humidity as to turn off evaporative cooling in order to solely detect the effect of background temperature. In Figure 2a, we exemplarily depict the influence of $T_{bg}$, for measurements at 19 °C, 26 °C (room temperature) and 40 °C, on the recorded apparent temperature of the FJ, $T_{cam}$. The leaf appears to be warmer if the background is warmer. This effect is the largest at the bottom of the leaf which is known to be the thinnest part. We then calculated the correction matrix of the FJ $\beta(x,z)$ using eq. (2) based on measurements at 30 different background temperatures in the range between 11 and 45 °C; exemplary fits are presented in Figure S1 of the SI.

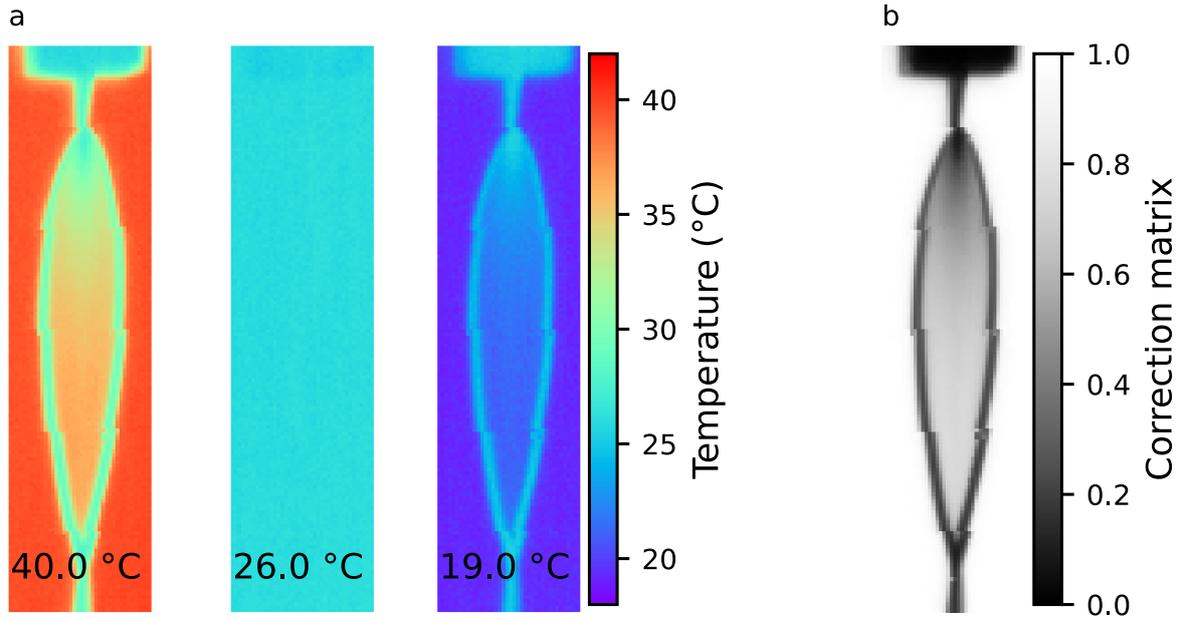

*Figure 2:* a) Exemplary infrared images of the liquid sheet measured at different background temperatures (40 °C, 26 °C (room temperature), and 19 °C). Here, the liquid jet is running at atmospheric pressure, such that the apparent temperature differences of the leaf solely result from the partial transmission of the background radiation. b) The correction matrix is calculated by applying equation (2) to measurements at 30 different background temperatures.

To elaborate on possible differences of $\beta(x,z)$ associated with water evaporative cooling (absent at 100% RH), we next consider analogous measurements under vacuum conditions. In this case, the effect of the background temperature on the measured temperature is not directly revealed due to the partial cancelation associated with water cooling and the transmittance of the background temperature. Nevertheless, the correction matrix for both cases is almost identical within the error bars; for minor differences see Figure S8 in the SI. Now, we can extract the temperature map of the FJ from the IR image of $T_{cam}(x,z)$ using eq. (3) and the correction matrix $\beta(x,z)$. Figure 3a, left hand, presents the IR $T_{cam}$-image recorded at 100% RH (corresponding to the middle subfigure in Fig. 2a shown on a different scale) at the right we show the image when measured in vacuum (0.01 mbar, corresponding to <0.1% RH). As expected, in the vacuum case the FJ appears to be colder, with the temperature $T_{cam}$ dropping by ~10 K (uncorrected) across the leaf. The extracted values of $T_{FJ}$ (eq. 3), and hence the resulting images are plotted in figure 3b. The temperature map $T_{FJ}(x,z)$ extracted such is independent on the background temperature, see Figs. S4 and S5. For the FJ running in a water atmosphere, we obtain a uniform temperature across the entire leaf (Fig. 3b left), at the value of the room temperature (26 °C). Our finding for atmospheric pressure disagrees with results reported in a recent Raman spectroscopy study.(*48*)

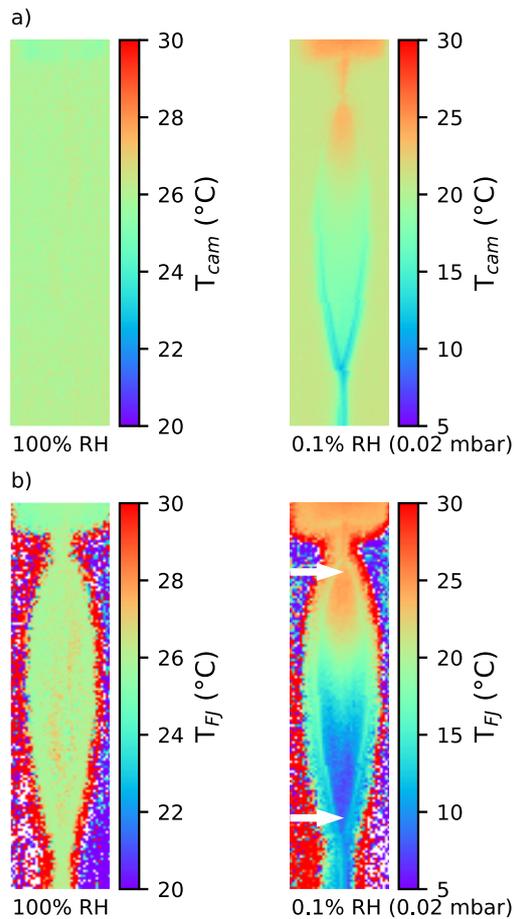

*Figure 3:* Temperature of liquid sheets at atmospheric pressure and in vacuum, respectively: a) Raw data ($T_{cam}$), recorded at 100% RH (left) and 0.1% RH (0.02 mbar) (right); b) The corrected data ($T_{FJ}$) reveal the true degree of evaporative cooling. The two white arrows mark

*the positions at the top and bottom part used for the calculation of the average cooling rate along the leaf. Note that the temperature ranges shown by color bars are different for the two measurements.*

In contrast, for the vacuum FJ (Figure 3b, right) evaporative cooling is seen to cause a temperature drop ($\Delta T$) of approximately ~15 K along the z-axis, measured from an approximate position where the leaf has formed up to the node where the second leaf evolves. In Fig. 3b (right), these points are marked with white arrows separated by $\Delta x$=2.7 mm distance.) With the flow rate along the z-axis in the leaf of $u_{FJ}$ ~21 m s$^{-1}$, the average cooling rate per unit of time, $K$, calculated using

$$K = \frac{\Delta T \cdot u_{FJ}}{\Delta x} \qquad (4),$$

yields $K = 1.13 \cdot 10^5$ K s$^{-1}$ (5.4 K mm$^{-1}$, for the FJ in vacuum). This is in agreement with the recent point-by-point measurements of cooling rates based on Raman spectroscopy from a FJ of almost identical jet parameters (flow and size).(*48*) The vacuum-jet temperature map also reveals changes along the x-axis (shorter dimension), with lower temperatures closer to the rim. This is consistent with the velocity maps recorded by Choo and Kang,(*55*) showing that the FJ flows faster in the middle. Hence, closer to the rim there is more time for evaporation and consequently lower temperatures are observed there.

In Figure 4 we evaluate the cooling rate as a function of % RH. The experimental procedure has been described in the Methods section. The % RH shown in the top axis is calculated from the average pressure in the chamber. Once pressures around 1 mbar (RH < 5%) are reached (upon strong pumping), cooling rates on the order of 10$^5$ K/s are observed. Higher pumping rates, leading to lower pressures in the chamber (which is usually favorable in, e.g., PES experiments (*1*)) have only a minor effect on the cooling rate. This means that data recorded in different experimental setups under vacuum conditions are comparable in terms of temperature, if the dimensions of the formed jet and its initial temperature (usually room temperature) are similar. At about 50% RH the cooling rate is halved compared to the maximum cooling rate.

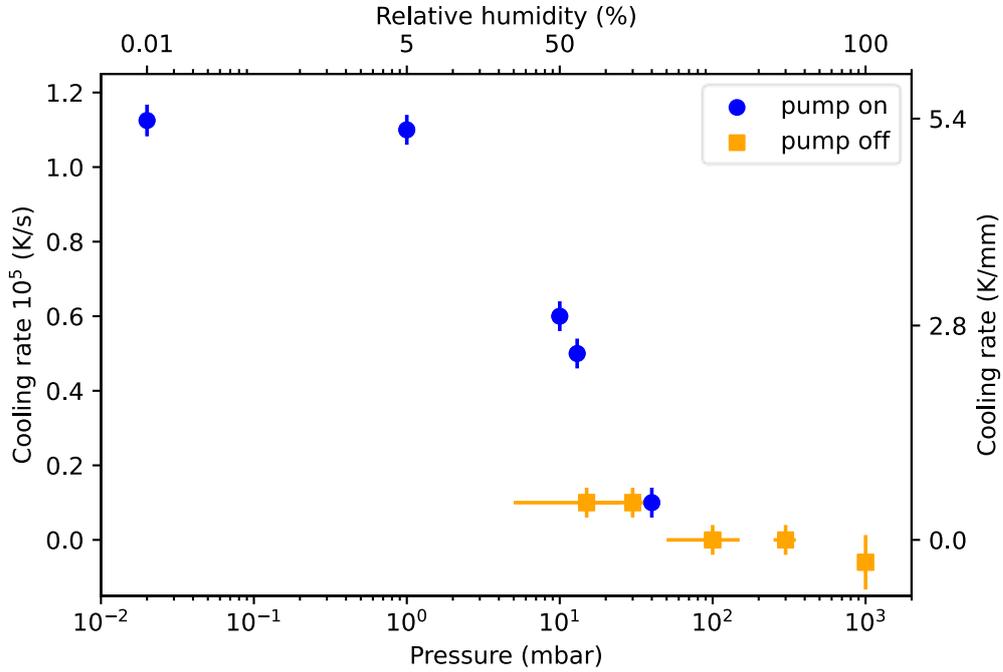

*Figure 4:* Average cooling rates (measured between top and bottom of leaf) of the flatjet as a function of background pressure. For the vacuum liquid jet, the cooling rate is ~1.13 $10^5$ K/s (or 5.4 K/mm, right x-axis). At 100% RH (top axis), the FJ temperature remains unchanged. Blue dots represent conditions with active pumping (via cold traps, roughing pumps or turbo pumps). The orange dots are measured without any pump being active (pump stopped after reaching that pressure, the x-error bars represent the uncertainty of the pressure). At about 50% RH (10 mbar) the cooling rate is halved compared to vacuum conditions.

In the remaining section we discuss the suitability of our experiment to determine spatially-resolved FJ thickness. Assuming that the calculated correction matrix $\beta(x,z)$ (eq. (2) and Figure 2b) primarily originates from the transmittance of IR radiation through the FJ, described by a spatially dependent transmission coefficient $\tau(x,z)$ (meaning $\beta(x,z) \approx \tau(x,z)$), we can extract information on the absolute thickness of the FJ, $d(x,z)$, by applying Lambert-Beer's law ($\tau = 10^{-\alpha \cdot c \cdot d}$). Here, $c$ is the molar concentration of water (55.5 mol/L). The molar absorption coefficient does depend on the wavelength, $\alpha(\lambda)$. In the case of our IR camera, we have to consider the wavelength range of 8–14 µm over which the signal intensity is integrated. This implies that we need to determine an average coefficient, $\alpha^*$. As detailed in the SI (Fig. S10) the molar absorption coefficient of water in the wavelength range of relevance is well documented. (*54, 56, 57*) It exhibits a steep decrease between 10 and 14 µm, and then stays rather flat within our region of interest, and importantly this spectral region is free of any sharp water features. Rather than determining α* directly from the grey-shaded region of Figure S10, we consider the ideal Planck spectrum of black body emission at 300 K, reproduced in the inset of Figure S11. Although, strictly speaking this spectrum may not exactly correspond to the actual (unknown) spectrum from room-temperature water it will be a very good approximation, and in addition potential small energy shifts would be irrelevant due to the signal integration over the wavelength detection window. We can then convolve the black-body emission spectrum with α(λ) to obtain the modified blue curve in Figure S11, which yields α* = 10.2 M$^{-1}$cm$^{-1}$. Considering the various molar absorption spectra reported in the literature we determine an error of the α* value of less than 5% (*54, 56, 57*). With the assumption $\beta(x,z) \approx \tau(x,z)$, and inserting the known value of *c* we can calculate the 2D image of the thickness of the FJ. (*56*) The obtained image is presented in Figure 5a, and a cut through the center line is shown in Figure 5b (black dots). The thickness at the center of the leaf is found to be 3.1 ± 0.4 µm; marked in the figure. (*8*) We note that Lambert-Beer´s law has some limitations though, for instance when applied to pure solvents, or in the case of chemical interactions at high concentrations; also surface effects are not captured.(*58, 59*) Applicability to liquid water has thus to be taken with care. Specifically, internal reflections at the FJ surface can lead to a higher measured transmittance than calculated with the Lambert-Beer law, such that the thickness in this case tends to be overestimated (also referred as effective thickness).(*58*) As the IR emission coefficient is approximately 0.95, we expect deviations due to internal reflection to be smaller than 10% (due to the assumption β≈τ). In order to estimate the importance of this effect based on experiment, we performed analogous measurements and determined a correction matrix (eq. 2) from a well characterized reported reference FJ produced with a chip nozzle(*17*), operated under conditions (impinging mode and 3.5 mL min$^{-1}$ flow rate) comparable to ref(*20*). We again used Lambert-Beer´s law to calculate the thickness of that leaf and obtained a thickness of 2.0±0.2 µm in the center of the leaf (see SI, Fig. S12) which is in very good agreement with the reported value of 2.1±0.1 µm determined by white light interferometry.(*20*)

The position-dependent thickness along the z-axis of the leaf can be fitted based on the Hasson-Peck model ($t = C1/z+C2$, blue line, with C1=1.91 and C2=2.29), (*31*) with the results presented in Figure 5. This model is known to less accurately describe the region near the rim where we indeed observe larger disagreement between experiment and fit. Figure 5b also shows the corresponding position-dependent temperatures (red curve), and it is seen that the thickness decreases faster than the temperature along the direction of flow (z-axis).

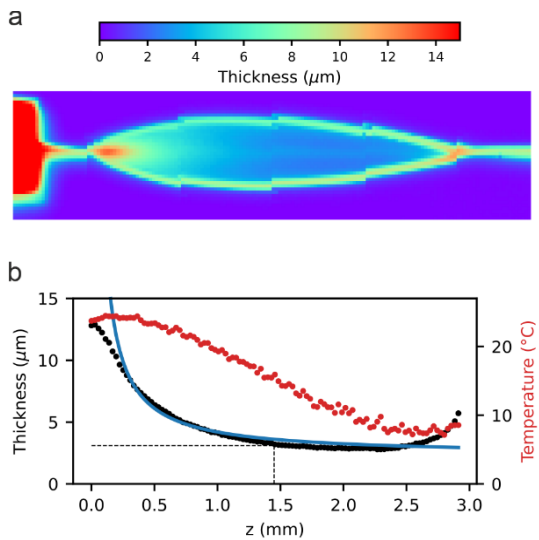

*Figure 5*: *a) Thickness of the flatjet calculated from infra-red camera images by applying the Lambert-Beer law. b) Thickness (black dots) along the z-axis of the leaf. The thickness is 3.1±0.4 µm in the center of the leaf. The blue line represents a simple fit applying the Hasson-Peck model. The temperature at those points is shown in red (right axis).*

In the last part of this communication, we consider the theoretical dependence of the thickness of the leaf as a function of evaporative cooling. The bulk of the liquid is cooled by evaporation of the surface molecules. Above we have determined a temperature drop of $\Delta T$ ~15 K across the length of the leaf under vacuum conditions (0.01 mbar or lower). We can then estimate which fraction of the liquid $f$ (average mass loss) has to be evaporated by comparing the enthalpy of evaporation $H_v$ and the drop of temperature $\Delta T$.(*60*) A similar balance equation was also used to calculate the fraction of ice initially formed from the supercooled liquid:(*61*)

$$f \cdot \Delta H_v = (1-f) \cdot c_p \cdot \Delta T \quad (5)$$

Here, $c_p$ is the molar heat capacity and $H_v$ the enthalpy of evaporation.(*62*) Solving eq. (8) for $x$ and using literature values for the heat capacity and enthalpy of evaporation results in:

$$f = \frac{c_p \cdot \Delta T}{\Delta H_v - c_p \Delta T} \approx 0.03 \quad (6)$$

That means that about 3% of the liquid evaporates based on the measured temperature differences across the leaf. This is in good agreement with the measured mass losses (3 to 5%) in other FJ studies.(*8, 26*) However, this is less than the accuracy of our thickness calculation, and hence the decrease of the thickness due to evaporation cannot be measured directly with the IR camera.

**Conclusion**

In conclusion, we have demonstrated that it is viable to measure and map the temperature and thickness profiles of partially transmissive thin liquid water sheets using an IR camera. Our main accomplishment is that this mapping is done in a single measurement, providing a 2D temperature image of the entire FJ surface, with a spatial resolution of a few tens of micrometers. For water FJs in vacuum, the inferred cooling rates are in very good agreement with measurements by Raman spectroscopy. As expected, no cooling is observed for the flat jet at atmospheric conditions. Our results also show that for the comparison of data from various laboratories, the exact knowledge of background pressure is not important as long as it is in the sub-mbar regime. On the other hand, flowrate, surface point of measurement and temperature of the liquid upon injection should be known. Our experimental setup and protocol are applicable to flat jets of sizes different than in the present study as long as the resolution of the IR camera used is sufficient to resolve the dimensions of the jet. For future experiments on liquid jets requiring an accurate knowledge of the local temperature, we recommend to use FJ instead of cylindrical LJ. Most important for future works is the ability to monitor temperature distributions, by an instantaneous 2D image, during chemical reactions occurring at the (aqueous) solution surface. This includes gas–liquid phase chemical reactions, or access of thermodynamic quantities associated with, e.g., temperature and/or pH-dependent solute molecular dissociation at the solution–vacuum interface.

**Author contributions:**

T.B. and B.W. planned the experiment. T.B. and H.H. performed the measurements. T.B. analyzed the data. T.B. wrote the manuscript with feedback of all authors.

**Supplementary information:**

Additional information on calibration and a simple finite element model are presented in the SI. Python scripts for image correction are available upon request.

**Acknowledgments**

The authors thank the FHI mechanical Workshop for constructing and maintaining the flatjet delivery system. T. B. thanks for discussions with S. E. Bradforth, H. C. Schewe, C. Richter and D. M. Stemer. T. B., U. H. and B. W. acknowledge funding from the European Research Council (ERC) under the European Union's Horizon 2020 research and investigation programme (Grant Agreement No. 883759-AQUACHIRAL).

# References

1. B. Winter, M. Faubel, Photoemission from Liquid Aqueous Solutions. *Chemical Reviews* **106**, 1176-1211 (2006).
2. T. Fransson *et al.*, X-ray and Electron Spectroscopy of Water. *Chemical Reviews* **116**, 7551-7569 (2016).
3. R. Dupuy *et al.*, Core level photoelectron spectroscopy of heterogeneous reactions at liquid–vapor interfaces: Current status, challenges, and prospects. *The Journal of Chemical Physics* **154**, 060901 (2021).
4. R. Signorell, B. Winter, Photoionization of the aqueous phase: clusters, droplets and liquid jets. *Physical Chemistry Chemical Physics* **24**, 13438-13460 (2022).
5. H. C. Schewe *et al.*, Photoelectron Spectroscopy of Benzene in the Liquid Phase and Dissolved in Liquid Ammonia. *The Journal of Physical Chemistry B* **126**, 229-238 (2022).
6. A. Hans *et al.*, Optical Fluorescence Detected from X-ray Irradiated Liquid Water. *The Journal of Physical Chemistry B* **121**, 2326-2330 (2017).
7. R. Golnak *et al.*, Joint Analysis of Radiative and Non-Radiative Electronic Relaxation Upon X-ray Irradiation of Transition Metal Aqueous Solutions. *Scientific Reports* **6**, 24659 (2016).
8. M. Ekimova, W. Quevedo, M. Faubel, P. Wernet, E. T. J. Nibbering, A liquid flatjet system for solution phase soft-x-ray spectroscopy. *Structural Dynamics* **2**, 054301 (2015).
9. F. Gel'mukhanov, M. Odelius, S. P. Polyutov, A. Föhlisch, V. Kimberg, Dynamics of resonant x-ray and Auger scattering. *Reviews of Modern Physics* **93**, 035001 (2021).
10. G. Taylor, Formation of Thin Flat Sheets of Water. *Proc R Soc Lon Ser-A* **259**, 1-& (1960).
11. A. Watanabe, H. Saito, Y. Ishida, M. Nakamoto, T. Yajima, A New Nozzle Producing Ultrathin Liquid Sheets for Femtosecond Pulse Dye-Lasers. *Opt Commun* **71**, 301-304 (1989).
12. J. W. M. Bush, A. E. Hasha, On the collision of laminar jets: fluid chains and fishbones. *Journal of Fluid Mechanics* **511**, 285-310 (2004).
13. G. Galinis *et al.*, Micrometer-thickness liquid sheet jets flowing in vacuum. *Review of Scientific Instruments* **88**, (2017).
14. M. Fondell *et al.*, Time-resolved soft X-ray absorption spectroscopy in transmission mode on liquids at MHz repetition rates. *Struct Dynam-Us* **4**, (2017).
15. B. Ha, D. P. DePonte, J. G. Santiago, Device design and flow scaling for liquid sheet jets. *Physical Review Fluids* **3**, 114202 (2018).
16. T. Debnath, M. S. B. Yusof, P. J. Low, Z. H. Loh, Ultrafast structural rearrangement dynamics induced by the photodetachment of phenoxide in aqueous solution. *Nature Communications* **10**, (2019).
17. J. D. Koralek *et al.*, Generation and characterization of ultrathin free-flowing liquid sheets (vol 1, 1353, 2018). *Nature Communications* **10**, (2019).
18. S. Menzi *et al.*, Generation and simple characterization of flat, liquid jets. *Review of Scientific Instruments* **91**, (2020).
19. C. B. Curry *et al.*, Cryogenic Liquid Jets for High Repetition Rate Discovery Science. *Jove-J Vis Exp*, (2020).
20. Z. H. Loh *et al.*, Observation of the fastest chemical processes in the radiolysis of water. *Science* **367**, 179-+ (2020).
21. J. Yang *et al.*, Direct observation of ultrafast hydrogen bond strengthening in liquid water. *Nature* **596**, 531-+ (2021).
22. M. S. B. Yusof, J. X. Siow, N. C. Yang, W. X. Chan, Z. H. Loh, Spectroscopic observation and ultrafast coherent vibrational dynamics of the aqueous phenylalanine radical. *Physical Chemistry Chemical Physics* **24**, 2800-2812 (2022).
23. C. J. Crissman *et al.*, Sub-micron thick liquid sheets produced by isotropically etched glass nozzles. *Lab Chip* **22**, 1365-1373 (2022).
24. D. J. Hoffman *et al.*, Liquid Heterostructures: Generation of Liquid–Liquid Interfaces in Free-Flowing Liquid Sheets. *Langmuir* **38**, 12822-12832 (2022).
25. J. C. T. Barnard *et al.*, Delivery of stable ultra-thin liquid sheets in vacuum for biochemical spectroscopy. *Frontiers in Molecular Biosciences* **9**, (2022).
26. D. J. Hoffman *et al.*, Microfluidic liquid sheets as large-area targets for high repetition XFELs. *Frontiers in Molecular Biosciences* **9**, (2022).
27. F. Treffert *et al.*, High-repetition-rate, multi-MeV deuteron acceleration from converging heavy water microjets at laser intensities of $10^{21}$ W/cm2. *Applied Physics Letters* **121**, 074104 (2022).
28. C. Lee *et al.*, Evaporation and Molecular Beam Scattering from a Flat Liquid Jet. *The Journal of Physical Chemistry A* **126**, 3373-3383 (2022).
29. R. Dupuy *et al.*, Ångstrom depth resolution with chemical specificity at the liquid-vapor interface. *Phys. Rev. Lett.*, (2023).
30. Y. J. Choo, B. S. Kang, The effect of jet velocity profile on the characteristics of thickness and velocity of the liquid sheet formed by two impinging jets. *Phys Fluids* **19**, (2007).
31. D. Hasson, R. E. Peck, Thickness distribution in a sheet formed by impinging jets. *Aiche Journal* **10**, 752-754 (1964).
32. Y. J. Choo, B. S. Kang, Parametric study on impinging-jet liquid sheet thickness distribution using an interferometric method. *Exp Fluids* **31**, 56-62 (2001).

33. S. Malerz *et al.*, A setup for studies of photoelectron circular dichroism from chiral molecules in aqueous solution. *Review of Scientific Instruments* **93**, 015101 (2022).
34. K. Modig, B. G. Pfrommer, B. Halle, Temperature-dependent hydrogen-bond geometry in liquid water. *Physical Review Letters* **90**, (2003).
35. P. Wernet *et al.*, The structure of the first coordination shell in liquid water. *Science* **304**, 995-999 (2004).
36. Y. X. Chen, N. Dupertuis, H. I. Okur, S. Roke, Temperature dependence of water-water and ion-water correlations in bulk water and electrolyte solutions probed by femtosecond elastic second harmonic scattering. *J Chem Phys* **148**, (2018).
37. C. M. Saak, I. Unger, G. Gopakumar, C. Caleman, O. Bjorneholm, Temperature Dependence of X-ray-Induced Auger Processes in Liquid Water. *J Phys Chem Lett* **11**, 2497-2501 (2020).
38. T. E. Gartner *et al.*, Anomalies and Local Structure of Liquid Water from Boiling to theSupercooled Regime as Predicted by the Many-Body MB-pol Mode. *J Phys Chem Lett* **13**, 3652-3658 (2022).
39. J. Meibohm, S. Schreck, P. Wernet, Temperature dependent soft x-ray absorption spectroscopy of liquids. *Review of Scientific Instruments* **85**, (2014).
40. M. Faubel, S. Schlemmer, J. P. Toennies, A molecular beam study of the evaporation of water from a liquid jet. *Zeitschrift für Physik D Atoms, Molecules and Clusters* **10**, 269-277 (1988).
41. T. Buttersack, V. C. Weiss, S. Bauerecker, Hypercooling Temperature of Water is about 100 K Higher than Calculated before. *J Phys Chem Lett* **9**, 471-475 (2018).
42. K. R. Wilson *et al.*, Investigation of volatile liquid surfaces by synchrotron x-ray spectroscopy of liquid microjets. *Review of Scientific Instruments* **75**, 725-736 (2004).
43. J. A. Sellberg *et al.*, Ultrafast X-ray probing of water structure below the homogeneous ice nucleation temperature. *Nature* **510**, 381-+ (2014).
44. C. Goy *et al.*, Shrinking of Rapidly Evaporating Water Microdroplets Reveals their Extreme Supercooling (vol 120, 015501, 2018). *Physical Review Letters* **120**, (2018).
45. J. D. Smith, C. D. Cappa, W. S. Drisdell, R. C. Cohen, R. J. Saykally, Raman thermometry measurements of free evaporation from liquid water droplets. *J Am Chem Soc* **128**, 12892-12898 (2006).
46. H. Suzuki, Y. Matsuzaki, A. Muraoka, M. Tachikawa, Raman spectroscopy of optically levitated supercooled water droplet. *J Chem Phys* **136**, (2012).
47. J. P. F. Nunes *et al.*, Liquid-phase mega-electron-volt ultrafast electron diffraction. *Structural Dynamics* **7**, 024301 (2020).
48. Y. P. Chang, Z. Yin, T. Balciunas, H. J. Worner, J. P. Wolf, Temperature measurements of liquid flat jets in vacuum. *Struct Dynam-Us* **9**, (2022).
49. S. Bauerecker, P. Ulbig, V. Buch, L. Vrbka, P. Jungwirth, Monitoring Ice Nucleation in Pure and Salty Water via High-Speed Imaging and Computer Simulations. *The Journal of Physical Chemistry C* **112**, 7631-7636 (2008).
50. H. C. Schewe *et al.*, Imaging of Chemical Kinetics at the Water–Water Interface in a Free-Flowing Liquid Flat-Jet. *Journal of the American Chemical Society* **144**, 7790-7795 (2022).
51. Y. J. Choo, B.-s. Kang, The velocity distribution of the liquid sheet formed by two low-speed impinging jets. *Physics of Fluids* **14**, 622-627 (2002).
52. Fluke. (2022).
53. . (Engineering ToolBox, 2003).
54. J. E. Bertie, Z. Lan, Infrared Intensities of Liquids XX: The Intensity of the OH Stretching Band of Liquid Water Revisited, and the Best Current Values of the Optical Constants of H2O(l) at 25°C between 15,000 and 1 cm−1. *Applied Spectroscopy* **50**, 1047-1057 (1996).
55. Y.-J. Choo, B.-S. Kang, The velocity distribution of the liquid sheet formed by two low-speed impinging jets. *Phys Fluids* **14**, 622-627 (2002).
56. S. Y. Venyaminov, F. G. Prendergast, Water (H2O and D2O) molar absorptivity in the 1000-4000 cm(-1) range and quantitative infrared spectroscopy of aqueous solutions. *Anal Biochem* **248**, 234-245 (1997).
57. J.-J. Max, C. Chapados, Isotope effects in liquid water by infrared spectroscopy. III. H2O and D2O spectra from 6000to0cm−1. *The Journal of Chemical Physics* **131**, 184505 (2009).
58. R. Brendel, The concept of effective film thickness for the determination of bond concentrations from IR spectra of weakly absorbing thin films on silicon. *Journal of Applied Physics* **69**, 7395-7399 (1991).
59. T. G. Mayerhöfer, H. Mutschke, J. Popp, Employing Theories Far beyond Their Limits—The Case of the (Boguer-)Beer–Lambert Law. *ChemPhysChem* **17**, 1948-1955 (2016).
60. M. Faubel, S. Schlemmer, J. P. Toennies, A Molecular-Beam Study of the Evaporation of Water from a Liquid Jet. *Z Phys D Atom Mol Cl* **10**, 269-277 (1988).
61. T. Buttersack, S. Bauerecker, Critical Radius of Supercooled Water Droplets: On the Transition toward Dendritic Freezing. *J Phys Chem B* **120**, 504-512 (2016).
62. J. R. Rumble, D. R. Lide, T. J. Bruno, *CRC Handbook of Chemistry and Physics: A Ready-reference Book of Chemical and Physical Data*. (CRC Press, 2019).